\documentclass[
    ,final            
  ]
  {aipproc}

\layoutstyle{6x9}

\newcommand{\Fg}{{F_2^\gamma}}
\newcommand{\Fp}{{F_2^p}}
\newcommand\qb{\bar q}
\newcommand{\GeV}{{\rm GeV}}
\newcommand{\PL}{{\rm PL}}
\newcommand{\HL}{{\rm HAD}}

\begin{document}
\newcommand{\ds}{\displaystyle}
\newcommand{\sss}{\scriptscriptstyle}
\def \eps{\varepsilon}
\def \deg{^{\circ}}
\newcommand{\shat}{\ensuremath{{\hat s}}}
\newcommand{\limit}{\ensuremath{_{\scriptscriptstyle \mathrm{lim}}}}
\newcommand{\lum}{\ensuremath{{\mathcal L}}}
\newcommand{\ubar}{\ensuremath{\bar{u}}}
\newcommand{\dbar}{\ensuremath{\bar{d}}}
\newcommand{\ebar}{\ensuremath{\bar{e}}}
\newcommand{\mubar}{\ensuremath{\bar\mu}}
\newcommand{\nubar}{\ensuremath{\bar\nu}}
\newcommand{\ra}{\rightarrow}
\def\prop{\sim}
\def\gapf{\ensuremath{f(\Delta\eta)}}
\def\lsim{\,\lower.25ex\hbox{$\scriptstyle\sim$}\kern-1.30ex%
\raise 0.55ex\hbox{$\scriptstyle <$}\,}
\newcommand {\pom} {I\!\!P}
\newcommand {\spom} {\mbox{\footnotesize{$\pom$}}}
\newcommand {\pomsub} {{\scriptscriptstyle \pom}}
\newcommand {\reg} {I\!\!R}
\newcommand {\regsub} {{\scriptscriptstyle \reg}}
\newcommand {\ppp} {\pom\pom\pom}
\newcommand {\ppr} {\pom\pom\reg}
\newcommand {\xpom} {x_{\pomsub}}
\newcommand {\apom} {\alpha_{\pomsub}}
\newcommand {\areg} {\alpha_{\regsub}}
\newcommand {\aprime} {\alpha^\prime_\pomsub}
\newcommand {\deta} {\Delta\eta}
\newcommand {\aveapom} {\bar{\alpha}_\pomsub}
\newcommand{\rfour}{\mbox{$r^{04}_{00}$}}
\newcommand{\rfive}{\mbox{$r5_{00}$}}
\newcommand{\rfivecomb}{\mbox{$r5_{00} + 2 r5_{11}$}}
\newcommand{\ronecomb}{\mbox{$r1_{00} + 2 r1_{11}$}}
\newcommand{\tprim}{\mbox{$t^\prime$}}
\newcommand{\tlc}{\mbox{$t$}}
\newcommand{\mxtwo}{M2_X}
\newcommand{\xbj}{x_{\mathrm{Bj}}}
\newcommand {\xl} {x_L}
\newcommand{\yjb}{y_{\scriptscriptstyle JB}}
\newcommand {\mxlps} {M_{X}^{\mathrm{LPS}}}
\newcommand{\gp}{\gamma p}
\newcommand{\gvp}{\gamma^* p}
\newcommand{\etamax}{\eta_{\mathrm{max}}}
\newcommand{\dargthree}{\xpom\,,\beta,\,Q^2}
\newcommand{\dargfour}{\xpom,\,t\,,\beta,\,Q^2}
\newcommand{\ftwod}{F_2^D}
\newcommand{\ftwodthree}{F_2^{D(3)}}
\newcommand{\ftwodfour}{F_2^{D(4)}}
\newcommand{\fldthree}{F_L^{D(3)}}
\newcommand{\fldfour}{F_L^{D(4)}}
\newcommand{\srdthree}{\sigma_r^{D(3)}}
\newcommand{\srdfour}{\sigma_r^{D(4)}}
\newcommand{\fonedthree}{F_1^{D(3)}}
\newcommand{\fonedfour}{F_1^{D(4)}}
\newcommand{\ftwodthreearg}{F_2^{D(3)}\,(\dargthree)}
\newcommand{\ftwodfourarg}{F_2^{D(4)}\,(\dargfour)}
\newcommand{\fldthreearg}{F_L^{D(3)}\,(\dargthree)}
\newcommand{\fldfourarg}{F_L^{D(4)}\,(\dargfour)}
\newcommand{\fonedthreearg}{F_1^{D(3)}\,(\dargthree)}
\newcommand{\fonedfourarg}{F_1^{D(4)}\,(\dargfour)}
\newcommand{\srdthreearg}{\sigma_r^{D(3)}\,(\dargthree)}
\newcommand{\srdfourarg}{\sigma_r^{D(4)}\,(\dargfour)}
\newcommand{\ftwopom}{F_2^{\pomsub}}
\newcommand{\flpom}{F_L^{\pomsub}}
\newcommand{\fonepom}{F_1^{\pomsub}}
\newcommand{\ftwopomarg}{\ftwopom\,(\beta,\,Q^2)}
\newcommand{\flpomarg}{\flpom\,(\beta,\,Q^2)}
\newcommand{\fonepomarg}{\fonepom\,(\beta,\,Q^2)}
\newcommand{\pomflux}{f_{\pomsub/p}}
\newcommand{\pomfluxargs}{\pomflux(\xpom,\,t)}
\newcommand{\pomfluxarg}{\pomflux(\xpom)}
\def\cts{\cos\theta^{\ast}}
\def\xgobs{x_\gamma^{\scriptscriptstyle \mathrm{OBS}}}
\def\xpobs{x_p^{\scriptscriptstyle \mathrm{OBS}}}
\def\xg{x_\gamma}
\def\ETJ{E_T^{\mathrm{jet}}}
\def\ETAJ{\eta^{\mathrm{jet}}}

\newcommand{\stat}{\mathrm{stat}}
\newcommand{\syst}{\mathrm{syst}}
\newcommand{\model}{\mathrm{model}}

\newcommand\units[1]{\,\mathrm{#1}} 
\newcommand\fig[1]{fig.\,\ref{fig:#1}} 
\newcommand\Fig[1]{Fig.\,\ref{fig:#1}}
\newcommand\tab[1]{table \ref{table:#1}}
\newcommand\Tab[1]{Table \ref{table:#1}}
\newcommand\qq[1]{Eq.\,(\ref{qq:#1})}
\newcommand\sect[1]{\S\ref{sect:#1}}

\def\ap#1#2#3   {{\em Ann. Phys. (NY)} {\bf#1} (#2) #3}   
\def\apj#1#2#3  {{\em Astrophys. J.} {\bf#1} (#2) #3} 
\def\apjl#1#2#3 {{\em Astrophys. J. Lett.} {\bf#1} (#2) #3}
\def\app#1#2#3  {{\em Acta. Phys. Pol.} {\bf#1} (#2) #3}
\def\ar#1#2#3   {{\em Ann. Rev. Nucl. Part. Sci.} {\bf#1} (#2) #3}
\def\cpc#1#2#3  {{\em Computer Phys. Comm.} {\bf#1} (#2) #3}
\def\epj#1#2#3  {{\em Eur. Phys. J.} {\bf#1} (#2) #3}
\def\err#1#2#3  {{\it Erratum} {\bf#1} (#2) #3}
\def\ib#1#2#3   {{\it ibid.} {\bf#1} (#2) #3}
\def\jmp#1#2#3  {{\em J. Math. Phys.} {\bf#1} (#2) #3}
\def\ijmp#1#2#3 {{\em Int. J. Mod. Phys.} {\bf#1} (#2) #3}
\def\jetp#1#2#3 {{\em JETP Lett.} {\bf#1} (#2) #3}
\def\jpg#1#2#3  {{\em J. Phys. G.} {\bf#1} (#2) #3}
\def\mpl#1#2#3  {{\em Mod. Phys. Lett.} {\bf#1} (#2) #3}
\def\nat#1#2#3  {{\em Nature (London)} {\bf#1} (#2) #3}
\def\nc#1#2#3   {{\em Nuovo Cim.} {\bf#1} (#2) #3}
\def\nim#1#2#3  {{\em Nucl. Instr. Meth.} {\bf#1} (#2) #3}
\def\np#1#2#3   {{\em Nucl. Phys.} {\bf#1} (#2) #3}
\def\pcps#1#2#3 {{\em Proc. Cam. Phil. Soc.} {\bf#1} (#2) #3}
\def\pl#1#2#3   {{\em Phys. Lett.} {\bf#1} (#2) #3}
\def\prep#1#2#3 {{\em Phys. Rep.} {\bf#1} (#2) #3}
\def\prev#1#2#3 {{\em Phys. Rev.} {\bf#1} (#2) #3}
\def\prl#1#2#3  {{\em Phys. Rev. Lett.} {\bf#1} (#2) #3}
\def\prs#1#2#3  {{\em Proc. Roy. Soc.} {\bf#1} (#2) #3}
\def\ptp#1#2#3  {{\em Prog. Th. Phys.} {\bf#1} (#2) #3}
\def\ps#1#2#3   {{\em Physica Scripta} {\bf#1} (#2) #3}
\def\rmp#1#2#3  {{\em Rev. Mod. Phys.} {\bf#1} (#2) #3}
\def\rpp#1#2#3  {{\em Rep. Prog. Phys.} {\bf#1} (#2) #3}
\def\sjnp#1#2#3 {{\em Sov. J. Nucl. Phys.} {\bf#1} (#2) #3}
\def\spj#1#2#3  {{\em Sov. Phys. JEPT} {\bf#1} (#2) #3}
\def\spu#1#2#3  {{\em Sov. Phys.-Usp.} {\bf#1} (#2) #3}
\def\zp#1#2#3   {{\em Zeit. Phys.} {\bf#1} (#2) #3}

\def\zn#1       {{\em Zeus Note} {\bf#1}}
\def\dn#1       {{\em Desy Note} {\bf#1}}
\def\hepph#1    {\mbox{arXiv:hep-ph/#1\/} }
\newcommand{\mx}{\ensuremath{M_{X}}}
\newcommand{\half}{\ensuremath{\frac{1}{2}}}
\newcommand{\gs}{\gamma^{\ast}}
\newcommand{\as}{\alpha_{S}}
\newcommand{\chipdf}{\frac{\chi^2}{\mbox{{\tiny d.o.f.}}}}
\newcommand{\der}[2]{\frac{d#1}{d#2}}
\newcommand{\pder}[2]{\frac{\partial#1}{\partial#2}}

\newcommand{\EPS}[3]{
  \noindent
  \begin{figure*}[htb]
  \begin{center}
  \epsfig{figure=#1.eps,#3}
  \caption{{\em #2}}
  \label{fig:#1}
  \end{center}
  \end{figure*}
}

\newcommand{\EPScp}[3]{
  \EPS{#1}{#2}{#3}
  \clearpage
}

\newcommand{\lpage}{\enlargethispage{\baselineskip}}
\newcommand{\spage}{\enlargethispage{-\baselineskip}}
\newcommand{\error}[1] {{\ensuremath{\pm #1}}}

\title{NLO PHOTON PARTON PARAMETRIZATION}


\author{Halina Abramowicz\footnote{also at Max Planck Institute, Munich, Germany, Alexander von Humboldt Research Award.}, Aharon Levy}{
  address={School of Physics and Astronomy, Raymond and Beverly Sackler Faculty of Exact Sciences, Tel Aviv University, Tel Aviv, Israel}
}

\author{Wojtek Slominski}{
  address={M. Smoluchowski Institute of Physics, Jagellonian University, 
Reymonta 4, 30-059, Cracow, Poland}
}


\begin{abstract}
  An NLO photon parton parametrization is presented based on the
  existing $F_2^\gamma$ measurements from $e^+e^-$ data and the
  low-$x$ proton structure function from $ep$ interactions. Also
  included in the extraction of the NLO parton distribution functions
  are the dijets data coming from $\gamma p \to j_1 + j_2 +X$.  The
  new parametrization is compared to other available NLO
  parametrizations.
\end{abstract}

\maketitle


\section{INTRODUCTION}

A new parametrization of the parton distributions in the photon is
extracted in next-to-leading order (NLO) of perturbative QCD. It
differs from other NLO parametrizations~\cite{GRV92,AFG,GS,GRS,CJK} in
that the data used in the fitting procedure include the expected
behaviour of $\Fg$ at low-$x$, as derived from $F_2^p$
measurements~\cite{f2p} under Gribov factorization
assumption~\cite{gribov-fact}, as suggested in~\cite{al-gribov} and,
in addition, the measurements of the dijet photoproduction cross
sections~\cite{dijets}.

\section{Gribov factorization}

It was suggested~\cite{al-gribov} that for low Bjorken $x$ ($x <$
0.01) one can use the relation based on Gribov
factorization~\cite{gribov-fact}, to find a simple relation between
$\Fg$ and $\Fp$. Gribov factorization relates the total $\gamma\gamma$
cross section to those of $\gamma p$ and $pp$. For low $x$ one can
thus obtain
\begin{equation}
\Fg(x,Q^2)=\Fp(x,Q^2)\frac{\sigma_{\gamma p}(W)}{\sigma_{pp}(W)}.
\end{equation}
Here $Q^2$ is the virtuality of the probing photon and $W$ is the
center of mass energy. Using the parameterization of Donnachie and
Landshoff~\cite{dl}, which gives a good representation of the data,
one obtains at large $W$
\begin{equation}
\Fg/\alpha = 0.43 \Fp,
\end{equation}
where $\alpha$ is the electromagnetic coupling constant. In extracting
parton distributions in the photon, this last relation allows the use
of the precise $\Fp$ data to constrain the low-$x$ region, where $\Fg$
data are very scarce.

\section{The parametrization}

Our parametrization of the initial parton distributions, defined at
$Q^2_0= 2\,\GeV^2$, aims at describing the experimental data below the charm
threshold. Thus we explicitly parametrize only the $u,d,s$ quarks and
the gluon. The $c,b$ and $t$ quarks are generated radiatively once
their respective thresholds are crossed.

All quark distributions in the photon are parametrized as a sum of 
point-like and hadron-like contributions,
\begin{equation}
f_q(x) = f_{\qb}(x) =
 e_q^2 {A^\PL} {x^2+(1-x)^2 \over 1- {B^\PL} \ln(1-x)}
   + f_q^\HL(x)
\,.
\end{equation}

Apart from the $e_q^2$ factor, the point-like contribution is the same
for all quarks. The hadron-like contribution is assumed to depend on
the quark mass only. For $u$ and $d$ quarks we parametrize it as
\begin{equation}
f_u^\HL(x) = f_d^\HL(x)
 = {A^\HL} x^{{B^\HL}} (1-x)^{{C^\HL}}
\,,
\end{equation}
and for the $s$ quark we fix it to be
\begin{equation}
f_s^\HL(x) = 0.3\, f_d^\HL(x)
\,.
\end{equation}

The gluons in the photon are assumed to have hadron-like behaviour
\begin{equation}
  f_G(x) = {A^\HL_G} x^{{B^\HL_G}} (1-x)^{{C^\HL_G}}
\,.
\end{equation}

As there are no data at $x$ close to 1 we fix 
$C^\HL = 1$ and $C^\HL_G = 3$ as suggested by counting
rules~\cite{Blankenbecler:1974tm,Farrar:yb} based on dimensional
arguments.  Thus we are left with 6 free parameters.

\section{The fit procedure and the data}

We use the DIS$_\gamma$ scheme to relate $\Fg$ to the parton
densities. We use the zero mass variable-flavor-number-scheme (VFNS)
for the DGLAP evolution of heavy flavor parton distribution functions
(pdfs). For the heavy quark contribution to $\Fg$ we adopt a
phenomenological parametrization as a weighted sum of the
Bethe-Heitler and pdf contributions~\cite{sal}. The weights are
defined so as to avoid double counting. The following masses of heavy
quarks were used: $m_c$ = 1.5 GeV, $m_b$ = 4.5 GeV and $m_t$ = 174
GeV.

For fitting the parameters we used all published data on the photon
structure function $\Fg$, from LEP, PETRA and TRISTAN~\cite{hawar}. We
also used the Gribov factorization relation in order to produce $\Fg$
'data' at low $x$ from the proton structure function data measured by
ZEUS~\cite{f2p}.  In addition the dijet photoproduction measurements
were taken from the ZEUS experiment~\cite{dijets}. All in all we used
164 points of $\Fg$ measurements coming from $e^+e^-$ reactions, 122
proton structure function data points from $ep$ interactions and 24
points of dijet photoproduction reactions.

\section{results}

\begin{figure}[h]
  \includegraphics[height=.6\textheight]{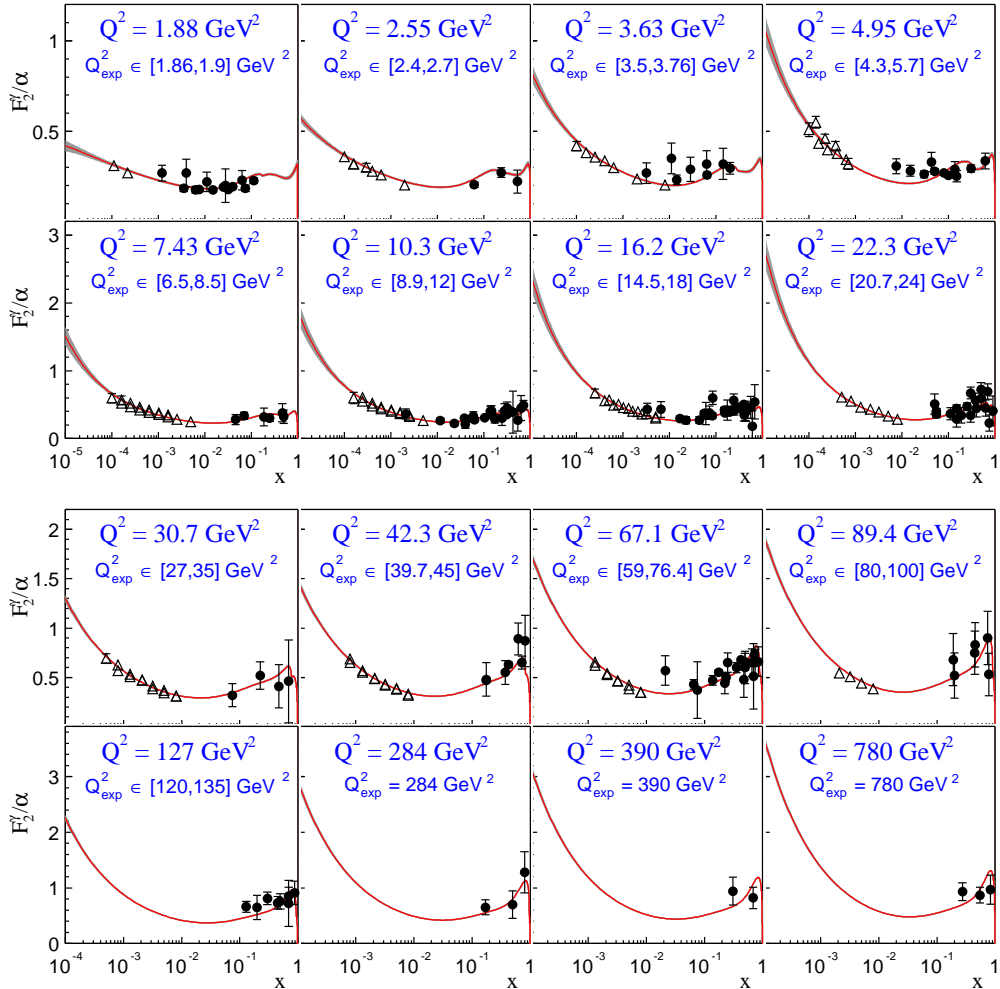}
  \caption{The SAL expectations for $\Fg(x,Q^2)$ as a function of $x$ at 
selected $Q^2$ values, as denoted in th figure. The plotted data 
(dots for $\Fg$ measured directly and triangles for $\Fg$ deduced from 
$\Fp$) are from the range $Q^2_{exp}$ presented in the figure}
\label{fig:f2g}
\end{figure}

The fit to the 286 structure function data points gave a value of 1.06
for the $\chi^2$ per degree of freedom. This increased to 1.63 when
the additional 24 dijets points were added. Nevertheless, it had only
a minor effect on the overall fit results and their errors. The best
fit expectations (denoted as the SAL parametrization), using all the
310 data points, are shown in figure~\ref{fig:f2g}, where $\Fg$ is
plotted as a function of $x$ in bins of $Q^2$.
The real $\Fg$ data and the ones deduced from $\Fp$ are shown with
different symbols. Note that wherever available, the two data sets
overlap within errors. To limit the number of plots without loss of
information, the data are shown within a range of $Q^2$, while the
corresponding curve is calculated for the average $Q^2$ of that bin.
The shaded error band is calculated according to the final error
matrix of the fitted parameters as returned by MINUIT.  The
uncertainty becomes smaller with increasing $Q^2$, due to the expected
loss of sensitivity to the initial pdf parametrization.

The dijet data gave a poor fit and did not help to constrain the
photon pdfs. The main reason is that the data are in a kinematical
region where the gluons in the proton dominate and thus may need to be
adjusted in order to get a better fit.

\section{Parton distributions}

\begin{figure}[h]
\includegraphics*[height=.3\textheight]{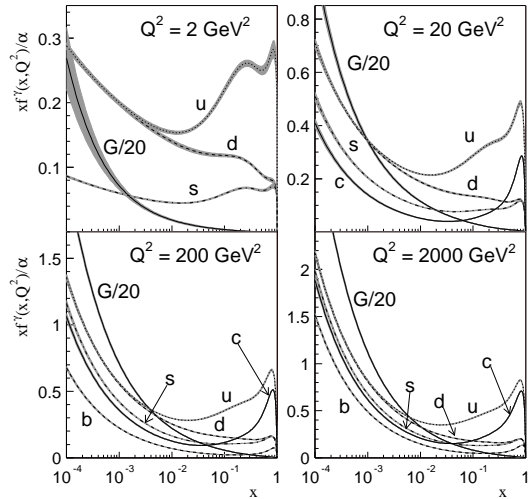}
  \caption{Parton distributions in the photon for different values of $Q^2$,
 as denoted in the figure.
}
\label{fig:pdfs}
\end{figure}

The SAL parton distributions in the photon are shown in
figure~\ref{fig:pdfs}.
The features to be noted are the behaviour of quarks at large $x$,
typical of the point-like contribution of the photon, and the
dominance of the gluon distribution at low $x$.

\begin{figure}[h]
\includegraphics*[height=.48\textheight]{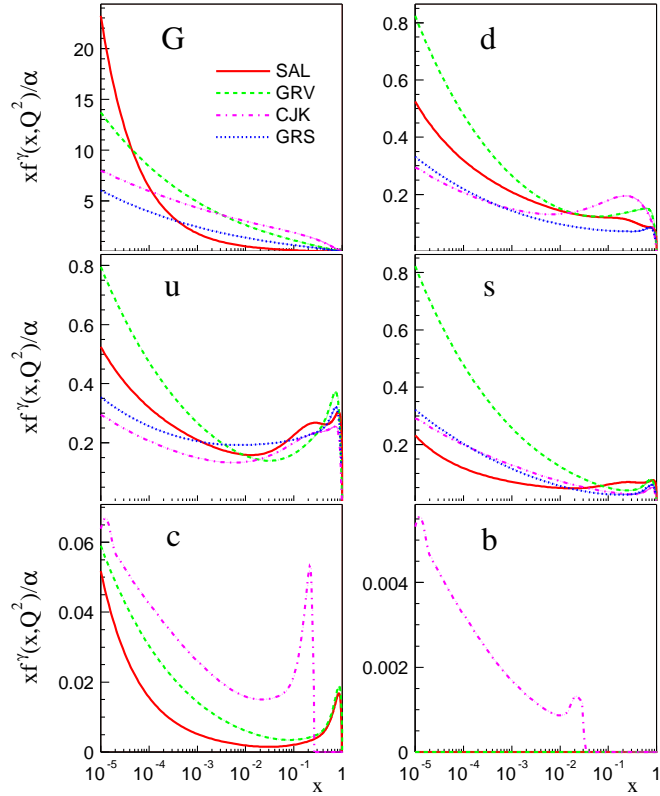}
  \caption{Comparison of SAL to other NLO parametrization at $Q^2$ = 2.5
 GeV$^2$.}
\label{fig:compare}
\end{figure}

The comparison of the SAL pdfs and the other available NLO
DIS$_\gamma$ photon parametrizations, GRV~\cite{GRV92},
GRS\footnote{This parametrization uses Fixed Flavor Number Scheme
  (FFNS), where only $u, d$ and $s$ pdfs exist.}~\cite{GRS}, and
CJK~\cite{CJK}, is shown in figure~\ref{fig:compare} for $Q^2$ = 2.5
GeV$^2$.
There are big differences between the various pdfs\footnote{A
  non-vanishing $b$-quark density at $Q^2$ = 2.5 GeV$^2$ is a feature
  of the CJK parametrization.}. They are especially pronounced for $x
< 10^{-3}$, where no $\Fg$ data are available and the result is
subject to additional theoretical assumptions. The SAL parametrization
has the lowest gluon distribution down to $x \sim 10^{-4}$, below
which value we observe a steep rise, steeper than other pdfs. At
higher $Q^2$, where the sensitivity to initial conditions is
diminished, there are still noticeable differences~\cite{sal}.


\begin{theacknowledgments}
This
  work was supported in part by the Israel Science Foundation (ISF).
\end{theacknowledgments}






\end{document}